\documentclass[12pt]{article}
\pdfoutput=1
\usepackage{tikz}
\usetikzlibrary{matrix,arrows,decorations.pathmorphing,decorations.pathreplacing}
\usepackage{jheppub}
\usepackage{amsmath,amssymb,euscript,array,mathrsfs,appendix,ctable,marvosym,calc,array}
\usepackage{arydshln}
\usepackage{todonotes}
\usepackage{graphicx}
\usepackage[normalem]{ulem}

\usepackage{blkarray}
\usepackage[normalem]{ulem}

\newcolumntype{C}[1]{>{\centering\arraybackslash$}m{#1}<{$}}
\newlength{\mycolwd}                                         
\newcommand\blank[1]{#1}
\renewcommand\blank[1]{}
\def\Buildrel#1\over#2\under#3{\mathrel{\mathop{\kern0pt
#2}\limits^{#1}_{#3}}}

\usepackage{color}
\definecolor{lightgray}{gray}{0.75}

\newcommand\greybox[1]{%
  \vskip\baselineskip%
  \par\noindent\colorbox{lightgray}{%
    \begin{minipage}{\textwidth}#1\end{minipage}%
  }%
  \vskip\baselineskip%
}

\def\FF{{\mathfrak F}}
\def\FFT{{\mathfrak F}^T}

\def\LL{{\mathscr L}}

\def\CF{{\cal F}}
\def\PP{{\mathbb P}}

\def\DG{d_{\mathfrak g}}

\def\DF{d_{\mathfrak f}}
\def\JJ{\mathscr{J}}

\def\mpsu{\mathfrak{psu}}

\def\SO{\text{SO}}

\def\mg{\mathfrak g}
\def\mf{\mathfrak f}
\def\mh{\mathfrak h}

\newcommand{\Tr}{\operatorname{Tr}}
\newcommand{\STr}{\operatorname{Str}}

\newcommand{\AD}{\operatorname{Ad}}

\def\B0{{\boldsymbol 0}}

\def\SU{\text{SU}}

\def\Dbarslash{\,\,{\raise.15ex\hbox{/}\mkern-12mu {\bar D}}}
\def\Dslash{\,\,{\raise.15ex\hbox{/}\mkern-12mu D}}
\def\delslash{\,\,{\raise.15ex\hbox{/}\mkern-9mu \partial}}
\def\delbarslash{\,\,{\raise.15ex\hbox{/}\mkern-9mu {\bar\partial}}}

\newcommand{\MAT}[1]{\begin{pmatrix} #1\end{pmatrix}}

\newcommand{\EQ}[1]{\begin{equation}\begin{split} #1
\end{split}\end{equation}}

\newcommand{\SP}[1]{\begin{equation}\begin{split} #1
\end{split}\end{equation}}

\newcommand{\FIG}[1]{\begin{figure}[ht]\begin{center} #1 \end{center}\end{figure}}

\title{An Integrable Deformation of the AdS$_{\boldsymbol 5}\boldsymbol\times \boldsymbol S^{\boldsymbol 5}$ Superstring}
\author[a]{Timothy J. Hollowood,}
\author[b]{J. Luis Miramontes}
\author[a,c]{and David M. Schmidtt}
\affiliation[a]{Department of Physics, Swansea University, Swansea, SA2 8PP, U.K.}
\affiliation[b]{Departamento de F\'\i sica de Part\'\i culas and IGFAE,
Universidad de Santiago de Compostela, 15782 Santiago de Compostela, Spain}
\affiliation[c]{Instituto de F\'\i sica Te\'orica IFT/UNESP, Rua Dr. Bento Teobaldo Ferraz 271, Bloco II, CEP 01140-070, Sa\~o Paulo-SP, Brasil}

\emailAdd{t.hollowood@swansea.ac.uk}
\emailAdd{jluis.miramontes@usc.es}
\emailAdd{david.schmidtt@gmail.com}

\abstract{The S-matrix on the world-sheet theory of the string in AdS$_5\times S^5$ has previously been shown to admit a deformation where the symmetry algebra is replaced by the associated quantum group. The case where $q$ is real has been identified as a particular deformation of the Green-Schwarz sigma model. An interpretation of the case with $q$ a root of unity has, until now, been lacking. We show that the Green-Schwarz sigma model admits a discrete deformation which can be viewed as a rather simple deformation of the $F/F_V$ gauged WZW model, where $F=\text{PSU}(2,2|4)$. The deformation parameter $q$ is then a $k$-th root of unity where $k$ is the level. The deformed theory has the same equations-of-motion as the Green-Schwarz sigma model but has a different symplectic structure.
We show that the resulting theory is integrable and has just the right amount of kappa-symmetries that appear as a remnant of the fermionic part of the original gauge symmetry.
This points to the existence of a fully consistent deformed string background.
}

\pgfdeclarelayer{top layer}
\pgfdeclarelayer{foreground layer}
\pgfdeclarelayer{background layer}
\pgfsetlayers{background layer,main,foreground layer,top layer}

\notoc
\begin{document}

\maketitle

\newpage

\section{Introduction}

The underlying integrability of the string propagating in AdS$_5\times S^5$ (reviewed in \cite{Beisert:2010jr,Arutyunov:2009ga}) can be traced to the fact that the 
target space of the string sigma model in the Green-Schwarz formalism is a very particular kind of quotient $F/G$ of a Lie supergroup $F=\text{PSU}(2,2|4)$ and  the ordinary Lie group $G=\text{Sp}(2,2)\times\text{Sp}(4)$, or equivalently $\SO(1,4)\times\SO(5)$. The bosonic part of the quotient $F/G$, which we denote $F_B/G$, is precisely AdS$_5\times S^5$ itself realised as a product of symmetric spaces $\SO(2,4)/\SO(1,4)\times\SO(6)/\SO(5)$. 

The quotient $F/G$ is known as a semi-symmetric space \cite{Serg} (see also \cite{Zarembo:2010sg}) and one of its crucial features is the existence of a $\mathbb Z_4$ automorphism of the Lie superalgebra  $\mf$ of $F$. These aspects are reviewed in appendix \ref{a1}.
This generalises the $\mathbb Z_2$ automorphism of an ordinary symmetric space. Under the automorphism $\mf$ decomposes into eigenspaces $\mf=\oplus_{i=0}^3\mf^{(i)}$ which is respected by the Lie superalgebra\footnote{The spaces $\mf^{(i)}$ are defined as linear combinations of generators with coefficients that are Grassmann even or odd depending on whether the generator is even or odd graded.}
\EQ{
[\mf^{(i)},\mf^{(j)}]\subset\mf^{(i+j\ \text{mod}\ 4)}\ .
\label{lgr}
}
The bosonic/fermionic parts of the superalgebra are precisely the even/odd graded components, and $\mf^{(0)}\equiv\mg$ is the Lie algebra of the ordinary Lie group $G$. The supertrace in the defining representation defines a bilinear form on the generators $T^a$:
\EQ{
\STr\big(T^aT^b\big)=\eta^{ab}\ .
\label{ip8}
}
We will always take a basis of generators which 
respects the $\mathbb Z_4$ grading and so $\eta^{ab}$ pairs generators of $\mf^{(1)}$ with $\mf^{(3)}$.

In a body of work \cite{Hoare:2011nd,Hoare:2011wr,Hoare:2012fc,Hoare:2013ysa}, deformations of the string world-sheet theory at the level of its exact S-matrix have been investigated. The idea is that one can write down a deformation of the S-matrix by deforming the symmetry structure of the world-sheet S-matrix which consists of two copies of the triply centrally extended Lie superalgebra $\mh=\mpsu(2|2)\ltimes{\mathbb R}^3$ by deforming it into the quantum group $U_q(\mh)^{\times2}$, where $q$ is the deformation parameter. The 2-body deformed S-matrix is built out of the $R$-matrix of $U_q(\mh)$ written down by Beisert and Koroteev \cite{Beisert:2008tw} in the schematic form
\EQ{
S(\theta_1,\theta_2)=\sigma(\theta_1,\theta_2)R_{U_q(\mh)}(\theta_1,\theta_2)\otimes
R_{U_q(\mh)}(\theta_1,\theta_2)\ .
}
Notice that since the gauge fixing of the world-sheet theory breaks relativistic invariance, the S-matrix elements are not functions of the rapidity difference of the two incoming particles. In fact, the analytically continued rapidity takes values on a torus rather than the cylinder familiar from a relativistic scattering theory. This feature complicates the analysis and there is a good deal of work involved in finding the dressing phase $\sigma(\theta_1,\theta_2)$ that ensures that all the S-matrix axioms are fulfilled \cite{Hoare:2011wr}. The S-matrix depends on two parameters: $q$, the parameter of the quantum group, and $g$ which is interpreted as the coupling constant of the world-sheet theory that in the context of the gauge/gravity correspondence is the 't~Hooft coupling of the dual gauge theory. 

There are two distinct choices one can make for $q$ which lead to rather different S-matrices that satisfy all the S-matrix axioms. Firstly, one can take $q$ real. In this case the deformation is also called the $\eta$-deformation where $q=e^{-\eta}$ \cite{Arutyunov:2013ega}. In this case the S-matrix is constructed using the vertex form of the $R$-matrix of the quantum group. 
This deformation has been identified at the Lagrangian level in \cite{Delduc:2013fga}. It  corresponds to a deformation of the target space background of the Green-Schwarz sigma model \cite{Delduc:2013qra} of the so-called Yang-Baxter type introduced by Klim\v{c}\'\i k \cite{Klimcik:2002zj} (see also \cite{Kawaguchi:2014qwa}). 
At the classical level the deformed theories have the same equations of motion 
but different symplectic structures \cite{Delduc:2013fga,Delduc:2013qra,Hollowood:2014fha}. The explicit background fields for this deformed theory were found in \cite{Arutyunov:2013ega} (see also \cite{Arutyunov:2012zt,Arutyunov:2012ai,Arutynov:2014ota,Arutyunov:2014cda,Delduc:2014kha,Hoare:2014pna} for other work on the deformed theory).

\FIG{
\begin{tikzpicture} [line width=1.5pt,inner sep=2mm,
place/.style={circle,draw=blue!50,fill=blue!20,thick}]
\begin{pgfonlayer}{foreground layer}
\node at (1.5,1.5) [place] (sm) {$S$}; 
\end{pgfonlayer}
\node at (0,0) (i1) {$i$};
\node at (3,0) (i2) {$j$};
\node at (0,3) (i3) {$k$};
\node at (3,3) (i4) {$l$};
\draw[->] (i1) -- (i4);
\draw[->] (i2) -- (i3);
\begin{scope}[xshift=6cm]
\begin{pgfonlayer}{foreground layer}
\node at (1.5,1.5) [place] (sm) {$S$}; 
\end{pgfonlayer}
\node at (0,0) (i1) {};
\node at (3,0) (i2) {};
\node at (0,3) (i3) {};
\node at (3,3) (i4) {};
\node at (0,1.5) (j1) {$a$};
\node at (1.5,0) (j2) {$b$};
\node at (3,1.5) (j3) {$c$};
\node at (1.5,3) (j4) {$d$};
\draw[->] (i1) -- (i4);
\draw[->] (i2) -- (i3);
\end{scope}
\end{tikzpicture}
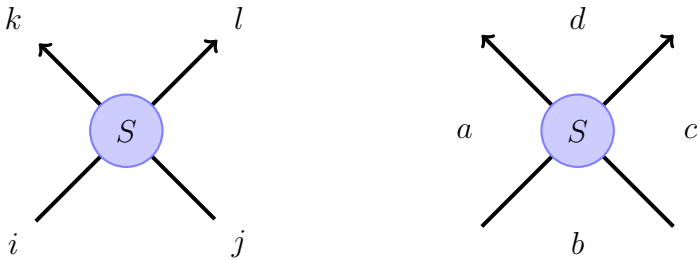
\caption{\small The vertex and IRF/RSOS labels for the 2-body S-matrix elements. For the case of $\SU(2)$ in the vertex picture, the particles are labelled by $i,j,k,l\in\{\pm\frac12\}$  the weights of the spin $\frac12$ representation of $\SU(2)$. The parameter $q$ must be real to satisfy Hermitian analyticity. In the IRF/RSOS picture on the right, one labels the vacua  $a,b,c,d\in\{0,\frac12,1,\frac32,\ldots\}$ which are spins of arbitrary irreducible representations of $\SU(2)$ with $|a-b|=|b-c|=|a-d|=|d-c|=\frac12$. The particles are then interpretated as kinks $K_{ab}+K_{bc}\to K_{ad}+K_{dc}$ interpolating between adjacent vacua. The parameter $q$ must be a complex phase to satisfy Hermitian analyticity. In the restricted case the spins are restricted to the finite set $\{0,\frac12,1,\ldots,\frac k2-1\}$, where $q=e^{i\pi/k}$.
\label{f1}
}}
When $q$ is a root of unity, the S-matrix takes the interaction-round-a-face (IRF) or restricted-solid-on-solid (RSOS) form, terms that are borrowed form the statistical mechanics application of quantum groups where the $R$-matrix elements play the role of Boltzmann weights. The difference between the vertex and IRF/RSOS form of the S-matrix elements is explained in figure \ref{f1} for the case $\SU(2)$. For the superstring S-matrix, there are effectively four copies of $\SU(2)$. The fact that the vertex form and IRF/RSOS forms are needed for $q$ real and root-of-unity, respectively, is governed by the need to satisfy the S-matrix axiom of Hermitian analyticity \cite{Hoare:2013ysa}.

The S-matrices for $q$ real and root-of-unity lead to rather different scattering theories. 
When one investigates the bound state structure using the bootstrap equations, in the former case the spectrum is infinite, whereas in the latter case the spectrum is finite \cite{Hoare:2012fc}. The latter is a direct consequence of the fact that when $q$ is a root-of-unity the set of vacua in the IRF picture is finite.

The goal of the present work is to present an explicit world-sheet world sheet theory that describes the $q$ deformed theories for $q$ a root of unity. The discussion generalizes that of \cite{us1} that addressed the same question for purely bosonic string theories on symmetric spaces.

In section \ref{s2} we briefly review some aspect of the Green-Schwarz sigma model for strings on AdS$_5\times S^5$. Section \ref{s3} describes how to deform the Green-Schwarz sigma model by exploiting its first order form and using an idea originally proposed to shed light on the global aspects of non-abelian T-duality. We then discuss integrability of the deformed theory and how to interpret it as a generalized sigma model with a non-trivial dilaton and fermionic WZ term. Section \ref{s3.5} contains a key part of the analysis because we show that the deformed theories admit local fermionic (kappa-)symmetries.
In section \ref{s6} we draw some conclusions.

\section{The Green-Schwarz Sigma Model and Integrability}\label{s2}

In this section we review some salient features of the Green-Schwarz theory of the superstring propagating in the AdS$_5\times S^5$ background written down by Metsaev and Tseytlin \cite{Metsaev:1998it}. We focus on this particular case but the approach should apply to other  AdS backgrounds.

The theory takes the form of a sigma model for the coset $F/G$ with a particular WZ term that is exact and so can be written in local form \cite{Berkovits:1999zq}. 
To this end we define a field $f(x,t)$ valued in the Lie supergroup $F$  and the current $J_\mu=f^{-1}\partial_\mu f$ in terms of which the action takes the form
\EQ{
S=-\frac{\kappa^2}{4\pi}\int d^2x\,\STr\Big[\sqrt{-h}\,h^{\mu\nu}J_\mu^{(2)}J^{(2)}_\nu+\varepsilon^{\mu\nu}J^{(1)}_\mu J^{(3)}_\nu\Big]\ .
}
In this formulation the 0-graded component of the current has been gauged away. Notice that the bosonic part of the theory is just a sigma model on a symmetry space $F_B/G$ and the fermionic fields couple via the WZ term. In most of what follows we work in conformal gauge $h_{\mu\nu}=e^\phi\eta_{\mu\nu}$ in which case we can write the action in terms of null components of the current as\footnote{In our notation $x^\pm=t\pm x$ and $\partial_\pm=(\partial_0\pm\partial_1)/2$ and so for vectors $A^\pm=A^0\pm A^1$ and $A_\pm=(A_0\pm A_1)/2$. The metric has components $h_{00}=-h_{11}=1$ and $\varepsilon^{01}=-\varepsilon^{10}=1$, so that $h^{+-}=h^{-+}=\varepsilon^{-+}=-\varepsilon^{+-}=2$.}
\EQ{
S=-\frac{\kappa^2}\pi\int d^2x\,\STr\Big[J_+^{(2)}J^{(2)}_--\frac12J^{(1)}_+J^{(3)}_-+\frac12J^{(1)}_-J^{(3)}_+\Big]\ .
\label{pss}
}

\subsection{Integrability}

The equations-of-motion of the sigma model, along with the Cartan-Maurer identity 
\EQ{
\partial_+J_--\partial_-J_++[J_+,J_-]=0\ ,
}
can be decomposed with respect to the $\mathbb Z_4$ gradation as the group of equations
\EQ{
&D_+ J^{(2)}_- +[J_+^{(1)},J_-^{(1)}]=0\ ,\\
&D_-J^{(2)}_+ + [J_-^{(3)},J_+^{(3)}]=0\ ,\\
&\partial_+J^{(0)}_--\partial_-J^{(0)}_++[J_+^{(0)},J_-^{(0)}]+[J_+^{(2)},
J_-^{(2)}]\\ &\hspace{2.8cm}+[J_+^{(3)},J_-^{(1)}]+[J_+^{(1)},J_-^{(3)}]=0\ ,\\
&D_+J_-^{(1)}-D_-J_+^{(1)}+[J_+^{(3)},J_-^{(2)}]=0\ ,\\
&D_+J_-^{(3)}-D_-J_+^{(3)}+[J_+^{(2)},J_-^{(1)}]=0\ ,\\
&[J_+^{(1)},J_-^{(2)}]=[J_+^{(2)},J_-^{(3)}]=0\ .
\label{eom4}
}
In the above, there is a $\mg$-valued connection $D_\pm\cdot  =[\partial_\pm+J_\pm^{(0)},\cdot]$.

The plethora of equations \eqref{eom4} can be written compactly in Lax form, which demonstrates integrability at the classical level \cite{Bena:2003wd,Alday:2005gi}:
\EQ{
[\partial_\mu+{\mathscr L}_\mu(z),\partial_\nu+{\mathscr L}_\nu(z)]=0\
\label{leq2}
}
with
\EQ{
{\mathscr L}_\pm(z)=J_\pm^{(0)}+z J_\pm^{(1)}+z^{\mp2} J_\pm^{(2)}+z^{-1} J_\pm^{(3)}\ ,
\label{rmm}
}
where $z$, the spectral parameter, is an arbitrary parameter.

\section{The Deformed Theory}\label{s3}

In order to define the deformation of the Green-Schwarz sigma model, we follow and suitably generalize the strategy of Sfetsos originally devised for the principal chiral model in \cite{Sfetsos:2013wia} (see also the earlier \cite{Rajeev:1988hq,Balog:1993es,Evans:1994hi,Sfetsos:1994vz,Polychronakos:2010hd}) and extended to the symmetric space sigma models in \cite{us1}.
The idea is to first write the sigma model in first order form which is essentially taking the sigma model in \eqref{pss} and viewing the currents, which we re-label $J_\mu\to A_\mu$, as fundamental. The fact that, actually, there exists a group-valued field $f$ such that $J_\mu=f^{-1}\partial_\mu f$ is then enforced by a Lagrange multiplier field $\nu$ in the Lie superalgebra $\mf$ that imposes the flatness of  $A_\mu$:
\EQ{
S=-\frac{\kappa^2}{\pi}\int d^2x\,\STr\Big[A_+^{(2)}A^{(2)}_--\frac12A^{(1)}_+A^{(3)}_-+\frac12A^{(1)}_-A^{(3)}_++\nu F_{+-}\Big]\ ,
\label{psr}
}
where
\EQ{
F_{+-}=\partial_+A_--\partial_-A_++[A_+,A_-]\ ,
}
is the single non-vanishing component of the curvature of $A_\mu$. 
The vanishing of $F_{+-}$ means that $A_\mu$ is pure gauge and that implicitly there exists a group valued field such that $A_\mu=f^{-1}\partial_\mu f$.

The action \eqref{psr} is the starting point to defining the T-dual of the Green-Schwarz sigma model with respect to the full $\text{PSU}(2,2|4)$ symmetry if one then integrates out $A_\mu$ instead of the Lagrange multiplier field $\nu$. 
Sfetsos \cite{Sfetsos:2013wia} proposes deforming the theory by replacing the Lagrange multiplier term in \eqref{psr} involving $\nu$ by the gauged WZW action for a group field $\CF$ valued here in the Lie supergroup $F$:
\EQ{
-\frac{\kappa^2}{\pi}\int d^2x\,\STr\big(\nu F_{+-}\big)\longrightarrow S_\text{gWZW}[\CF,A_\mu]\ .
}
Note that the gauge field is also valued in $\mathfrak{f}$ and so the WZW model is an $F/F_V$ gauged WZW theory for the supergroup $F$ gauged with respect to the anomaly free vector subgroup of $F_L\times F_R$. The intuition here is that in the limit when the level of the WZW model becomes large, $k\to\infty$, we can expand $\CF=1+\kappa^2\nu/k+\cdots$ in which case
\EQ{
S_\text{gWZW}[\CF=e^{\kappa^2\nu/k},A_\mu]\ \overset{k\to\infty}=\ 
-\frac{\kappa^2}{\pi}\int d^2x\,\STr\big(\nu F_{+-}\big)+\cdots\ .
}

Then, the proposal for the deformed theory is therefore
\EQ{
&S_\text{def}[\CF ,A_\mu]=S_\text{gWZW}[\CF ,A_\mu]-\frac {k}{\pi}\int d^2x\,\STr\,\big[
A_+\big(\Omega-1\big)A_-\big]\\
&=S_\text{gWZW}[\CF ,A_\mu]\\ &\qquad-\frac {k}{\pi}\int d^2x\,\STr\,\big[
(s_1-1)A_+^{(3)}A_-^{(1)}+(s_2-1)A_+^{(2)}A_-^{(2)}+(s_3-1)A_+^{(1)}A_-^{(3)}\big]
\label{dWZW}
}
where
\EQ{
\Omega=\PP^{(0)}+s_1\mathbb P^{(1)}+s_2\mathbb P^{(2)}+s_3\mathbb P^{(3)}\ ,
\label{dom}
}
and the action of the gauged WZW action is \cite{Karabali:1988au,Gawedzki:1988hq,Karabali:1989dk}
\EQ{
S_\text{gWZW}[\CF ,A_\mu]&=-\frac k{2\pi}\int d^2x\STr\Big[
\CF ^{-1}\partial_+\CF \,\CF ^{-1}\partial_-\CF +2A_+\partial_-\CF \CF ^{-1}\\ &~~~~~~~~~
-2A_-\CF ^{-1}\partial_+\CF -2\CF ^{-1}A_+\CF  A_-+2A_+A_-\Big]
\\ &~~~~~~~~~+\frac k{12\pi}\int d^3x\,\epsilon^{abc}\Tr\,\Big[\CF ^{-1}\partial_a\CF \,
\CF ^{-1}\partial_b\CF \,\CF ^{-1}\partial_c\CF \Big]\ .
\label{gWZW}
}

Notice that $\Omega-1$ vanishes on $\mf^{(0)}$ manifesting the fact that 
the deformation explicitly breaks the $F$ gauge symmetry to the subgroup $G$, the group in the denominator of the semi-symmetric space. So, strictly speaking, only the component $A^{(0)}_\mu$ is a true gauge field even though we will often refer to the full $A_\mu$ as a ``gauge field".

The couplings $s_i$ have yet to be determined but, in order to recover the sigma model in the limit $k\to\infty$, we require that they have the asymptotic behaviour
\EQ{
s_1\longrightarrow1+\frac{\kappa^2}{2k}+\cdots\ ,\qquad
s_2\longrightarrow1+\frac{\kappa^2}{k}+\cdots\ ,\qquad
s_3\longrightarrow1-\frac{\kappa^2}{2k}+\cdots\ .
}

\subsection{Integrability}

The equations-of-motion for the gauge field give the constraints
\EQ{
\JJ_+=-\frac k{2\pi}\big(\Omega^TA_+-A_-\big)\ ,\qquad\JJ_-=-\frac k{2\pi}\big(\Omega A_--A_+\big)\ ,
\label{con}
}
where we have defined the usual currents of the gauged WZW model,
\EQ{
\JJ_+&=-\frac k{2\pi}\big(\CF ^{-1}\partial_+\CF +\CF ^{-1}A_+\CF -A_-\big)\ ,\\
\JJ_-&=\frac k{2\pi}\big(\partial_-\CF \CF ^{-1}-\CF A_-\CF ^{-1}+A_+\big)\ .
\label{curr}
}
In \eqref{con} we have defined the transpose 
\EQ{
\Omega^T=\PP^{(0)}+s_3\mathbb P^{(1)}+s_2\mathbb P^{(2)}+s_1\mathbb P^{(3)}\ ,
\label{dot}
}
so that $\STr(A_+\Omega A_-)=\STr(A_-\Omega^T A_+)$. Note that because of the behaviour of the inner product of the Lie superalgebra \eqref{ip8}, the couplings $s_1$ and $s_3$ swap over.

The equation-of-motion of the group field $\CF$ can be written either as
\EQ{
\big[\partial_++\CF ^{-1}\partial_+\CF +\CF ^{-1}A_+\CF ,\partial_-+A_-\big]=0\ ,
\label{sh2}
}
or, equivalently, by conjugating with $\CF$, as
\EQ{
\big[\partial_++A_+,\partial_--\partial_-\CF \CF ^{-1}+\CF  A_-\CF ^{-1}\big]=0\ .
\label{sh3}
}

The strategy is to fix the couplings $s_i$ by demanding that the deformed theory is integrable. More specifically, we will demand that the equations-of-motion can be written as a Lax equation of the same form of the sigma model \eqref{leq2}. If this is to be possible then it must be that the two equations \eqref{sh2} and \eqref{sh3} are equivalent to \eqref{leq2} with particular values of the spectral parameter $z_\pm$, respectively; i.e.
\EQ{
-\frac{2\pi}k\JJ_\pm+A_\mp=\LL_\pm(z_\pm)\ ,\qquad A_\pm=\LL_\pm(z_\mp)\,,
\label{hyw}
}
where $\LL_\pm(z)$ takes the form \eqref{rmm}. These conditions are very constraining
since they determine all the couplings
\EQ{
s_1=\frac{1}\lambda\ ,\qquad s_2=\frac1{\lambda^2}\ ,\qquad s_3=\lambda\ ,
\label{couplings}
}
and, therefore,
\EQ{
\Omega&=\PP^{(0)}+\frac{1}\lambda\PP^{(1)}+\frac{1}{\lambda^2}\PP^{(2)}+\lambda\PP^{(3)}\ ,\\
\Omega^T&=\PP^{(0)}+\lambda\PP^{(1)}+\frac{1}{\lambda^2}\PP^{(2)}+\frac{1}\lambda\PP^{(3)}\ ,
}
where we have defined $\lambda=z_+/z_-$. In addition, they require that all the constraints \eqref{con} must be imposed: as usual integrability is a stern mistress and there is no leeway. Without loss of generality, we can take
\EQ{
z_\pm=\lambda^{\pm1/2}\ .
}

The fact that the equations-of-motion of the deformed WZW theory are equal to the Lax equation \eqref{leq2} for two distinct values of $z$ does not, on the face of it, seem to be strong enough to prove equivalence to the Lax equation \eqref{leq2} which holds for arbitrary $z$. The reason why it is sufficient is that the Lax equation has terms with powers $z^p$ ranging from $p=-3$ to $p=3$. However, for each projection of \eqref{leq2} on $\mf^{(i)}$ there are at most two independent equations, e.g.~for $\mf^{(3)}$ these are equations of order $z^{-1}$ and $z^3$. Consequently it follows that if the  Lax equation holds at two distinct values of $z$ then it holds for arbitrary $z$.

The relation between the WZW gauge field and the current $J_\mu$ of the original Green-Schwarz sigma model follows from \eqref{rmm} and \eqref{hyw} as
\EQ{
A_\pm^{(0)}&=J_\pm^{(0)}\ ,\qquad~\,  A_\pm^{(1)}=\lambda^{\mp1/2}J_\pm^{(1)}\ ,\\
A_\pm^{(2)}&=\lambda J_\pm^{(2)}\ ,\qquad A_\pm^{(3)}=\lambda^{\pm1/2}J_\pm^{(3)}\ .
\label{rel}
}
Using the constraints \eqref{con}, we can write this dictionary in terms of the WZW currents as
\EQ{
\JJ^{(0)}_\pm&=\mp\frac k{2\pi}J_1^{(0)}\ ,\hspace{4cm}
\JJ_\pm^{(1)}=\mp\frac k{2\pi}
\lambda^{\pm1/2} J_1^{(1)}\ ,\\ 
\JJ_\pm^{(2)}&=\mp\frac k{2\pi}\left(\lambda^{\mp1}J_+^{(2)}-\lambda^{\pm1}J^{(2)}_-\right)\ , \qquad
\JJ_\pm^{(3)}=\mp\frac k{2\pi}\lambda^{\mp1/2}J_1^{(3)}\  .
\label{jk3}
}

In the limit $\lambda\to1$, we identify
\EQ{
\frac{1}\lambda=1+\frac{\kappa^2}{2k}+\cdots\ ,
}
in which case, the deformation in \eqref{dWZW} becomes
\EQ{
-\frac {\kappa^2}\pi\int d^2x\,\STr\Big[A_+^{(2)}A_-^{(2)}-\frac12 A_+^{(1)}A_-^{(3)}+\frac12A_+^{(3)}A_-^{(1)}
\Big]\ ,
}
which shows that in this limit the theory returns to the Green-Schwarz sigma model \eqref{psr} with $J_\pm$ identified with $A_\pm$.

\subsection{An effective sigma model}\label{3.2}

It is possible to integrate out the gauge field $A_\mu$ to get an effective sigma model for the group-valued field $\CF$. One important thing to stress is that, for generic $\lambda$, the theory actually only has a (conventional) gauge symmetry $G$ so that only the component $A^{(0)}_\mu$ is a true gauge field. This also means that ultimately we should gauge fix this symmetry. However, it is a key part of our analysis that some of the broken gauge symmetry in the fermionic sector re-emerges and play the r\^ole of the all-important kappa-symmetries of a Green-Schwarz sigma model. The other important point is that, at the quantum level, the determinant that arises in integrating out $A_\mu$ gives rise to a dilaton on the world-sheet.

The deformed gauged WZW models that we are considering are similar to those considered by 
Tseytlin in \cite{Tseytlin:1993hm}. In that reference, there is a matrix
$Q=2-\Omega$; however, in the current case $\Omega$, and hence $Q$, is not symmetric ($\Omega\not=\Omega^T$) and thus, in addition to the fact that we are considering a supergroup, our theory lies outside the class considered by Tseytlin.

The equations-of-motion of the gauge field can be used to express the gauge field in terms of $\CF$: 
\EQ{
A_+&=-\big(\FFT-\Omega^T\big)^{-1}\CF^{-1}\partial_+\CF\ ,\\
A_-&=\big(1-\FFT\Omega\big)^{-1}\CF^{-1}\partial_-\CF\ ,
\label{icr}
}
where $\FF=\AD(\CF)$ and $\FFT=\AD(\CF^{-1})$, so that $\FF\FFT=1$,
and the resulting action is
\EQ{
S_\text{eff}=-\frac k{2\pi}\int d^2x\,\STr\Big[
\CF ^{-1}\partial_+\CF\big(1-2\big(1-\FFT\Omega\big)^{-1}\big)\CF ^{-1}\partial_-\CF\Big]
+S_\text{WZ}+S_\text{dil}\ .
\label{ger}
}
Here, $S_\text{WZ}$ is the Wess-Zumino term and, re-instating the world-sheet metric, the dilaton term is
\EQ{
S_\text{dil}=\frac1{4\pi}\int d^2x\,\sqrt{h}\ R^{(2)}\phi\ ,\qquad\phi=-\frac12\STr\log\big(\FF-\Omega\big)\ .
}

We can write the effective sigma model as a sum of a metric and $B$-field term:
\EQ{
S_\text{eff}=-\frac k{2\pi}\int d^2x\,\STr\Big[
\CF ^{-1}\partial_+\CF\big(G+B\big)\CF^{-1}\partial_-\CF\Big]+S_\text{dil}\ .
\label{ge3}
}
where
\EQ{
G&=1-\big(1-\FFT\Omega\big)^{-1}-\big(1-\Omega^T\FF\big)^{-1}\\[5pt]
&=\big(\FF-\Omega\big)^{-1}\big(\Omega\Omega^T-1\big)\big(\FFT-\Omega^T\big)^{-1}\\[5pt]
&=(1-\Omega^T\FF)^{-1} \big(\Omega^T\Omega-1)(1-\FF^T\Omega)^{-1}
\ ,
\label{metric}
}
and
\EQ{
B=B_0-\big(1-\FFT\Omega\big)^{-1}+\big(1-\Omega^T\FF\big)^{-1}\ ,
}
where $B_0$ indicates the contribution from the WZ term. 

Given the form of $\Omega$ and $\Omega^T$ in \eqref{dom} and \eqref{dot}, and taking into account~\eqref{couplings}, we see that 
\EQ{
\Omega\Omega^T-1=\Omega^T\Omega-1=\big(\lambda^{-4}-1\big){\mathbb P}^{(2)}\ ,
\label{xnn}
}
so that the $G$ term above defines a metric on the physical bosonic coordinates (after gauge fixing: see below). If we define the deformed frame
\EQ{
E^a=\STr\Big[T^a(1-\FFT\Omega)^{-1}\CF^{-1}d\CF\Big]\ ,
}
then the metric has components
\EQ{
g_{ij}=\big(\lambda^{-4}-1\big)\sum_{T^a\in\mf^{(2)}}\eta_{ab}E^a_iE^b_j\ .
}
Note that the Grassmann 1-forms $E^a$, $a=1,3$, only couple through the $B$-field term. This is a characteristic feature of Green-Schwarz sigma models, e.g.~\eqref{pss} and in more general background geometries \cite{Henneaux:1984mh,Grisaru:1985fv,Metsaev:1998it}. 
Later we will see that there are also local fermionic (kappa-) symmetries which means that the sigma model needs some suitable form of gauge fixing in the fermionic sector as well as the conventional gauge symmetry. All this motivates the interpretation of the target space as a consistent string background. 

If we temporarily focus on the bosonic sector by setting the fermions to zero, i.e~take $\CF\in F_B$, the bosonic subgroup of $F$, then $\Omega=\PP^{(0)}+\lambda^{-2}\PP^{(2)}$ which is now symmetric. In this case,
the sigma model lies in the class defined by Tseytlin \cite{Tseytlin:1993hm}. Since we have integrated out the gauge field we must impose a unitary gauge fixing on the field $\CF$ by choosing a suitable gauge choice. For the symmetric spaces involving orthogonal groups
$\text{AdS}_n=\SO(2,n-1)/\SO(1,n-1)$ and $S^n=\SO(n+1)/\SO(n)$, a suitable gauge slice is defined in \cite{Fradkin:1991ie,Grigoriev:2007bu,Grigoriev:2008jq}. In the latter case, the slice is defined in terms
of the $\SO(2)$ subgroups
\EQ{
g_m(\theta)=e^{\theta R_m}\ ,\qquad (R_m)_i{}^j=\delta_m{}^i\delta_{m+1}{}^j-\delta_m{}^j\delta_{m+1}^i\ ,
}
where $m=1,2,\ldots,n$ and $i,j=1,2\ldots,n+1$. The $\SO(n)$ subgroup corresponds to limiting $i,j=1,2,\ldots,n$. The slice takes the form
\EQ{
\CF=g_1(\theta_1)\cdots g_{n-1}(\theta_{n-1})g_{n}(2\theta_{n})
g_{n-1}(\theta_{n-1})\cdots g_1(\theta_1)\ .
}
It turns out that with this gauge slice, the WZ term $B_0$ and the full $B$-field vanish (for reasons that are similar to the situation in \cite{Grigoriev:2007bu}). The interpretation is that there is no NS flux in the spacetime geometry. Of course it would be very interesting to add in the fermions and extract the RR fluxes.
For the simplest case of $S^2$, the explicit metric is
\EQ{
g_{ij}=\frac1{1-\lambda^4}\left (
\begin{array}{*{2}{@{}C{\mycolwd}@{}}}
   (1+\lambda^4-2\lambda^2\cos2\theta_1)\cot^2\theta_2 & 2\lambda^2\sin2\theta_1\cot\theta_2\\
2\lambda^2\sin2\theta_1\cot\theta_2 & 1+\lambda^4+2\lambda^2\cos2\theta_1
\end{array}
\right )\ .
}

It is not surprising that the bosonic sigma model is not conformally invariant. Tseytlin \cite{Tseytlin:1993hm} computes the one-loop renormalization group flow of these theories and, in the present case, the running can be absorbed into the coupling $\lambda$ with\footnote{In \cite{Tseytlin:1993hm}, the running is defined in terms of 
a matrix $K=(\Omega+1)/(\Omega-1)\PP^{(2)}$ which has been restricted to lie orthogonal to the  gauge group (on which $\Omega=1$). In the present case, therefore, $K=(1+\lambda^2)/(1-\lambda^2)\PP^{(2)}$. The beta function is then $\mu\,dK_{rs}/d\mu=(8k)^{-1}c_2(F_B)(\delta_{rs}-K_{rt}K_{st})$ which gives the expression quoted here. See \cite{Sfetsos:2013wia,Sfetsos:2014jfa,Itsios:2014lca} for more recent and general calculations of this type.}
\EQ{
\mu\frac{d\lambda}{d\mu}=-\frac{c_2(F_B)}{8k}\lambda\ .
}

It would be interesting to confirm that the fermionic contributions to this one-loop beta function cancel the bosonic contributions along the lines of a similar calculation for the AdS$_5\times S^5$ Green-Schwarz sigma model in \cite{Zarembo:2010sg}. It is likely that the fact that $c_2(F)=0$ for $F=\text{PSU}(2,2|4)$ will be important in this regard.

\section{Symmetries}\label{s3.5}

Symmetries play a key r\^ole in this analysis because they ensure that the deformed theories have the correct number of degrees-of-freedom to be identified as a superstring world-sheet theory. 

The deformed theories have an obvious $G$ gauge symmetry under which, infinitesimally,
\EQ{
\delta\CF=[\epsilon^{(0)},\CF]\ ,\qquad \delta A_\mu=[\epsilon^{(0)},A_\mu]-\partial_\mu\epsilon^{(0)}\ ,
}
for $\epsilon^{(0)}\in\mf^{(0)}$.
Once this symmetry is fixed, na\"\i vely, the physical bosonic and fermionic 
dimension of configuration space is 
equal to $\DF^{(2)}$ and $\frac12(\DF^{(1)}+\DF^{(3)})$, respectively.\footnote{In counting the dimension of configuration space fermions effectively contribute $\frac12$.}
For $\text{PSU}(2,2|4)/\text{Sp}(2,2)\times\text{Sp}(4)$, assuming that $A_\pm^{(2)}$ are generic, we have
\EQ{
&\DG=20\ ,\quad\DF^{(1)}=\DF^{(3)}=16\ ,\quad\DF^{(2)}=10\ .
}
As it stands, therefore, there is no matching between the number of bosonic and fermionic  degrees-of-freedom as would be required to have a consistent string world-sheet theory.

However, we have missed the fact that the world-sheet has additional constraints in the form of the Virasoro conditions 
\EQ{
T_{\pm\pm}=0\ .
\label{vir}
} 
The components of the energy-momentum tensor can be obtained directly from the effective action~\eqref{ge3} which, once the world-sheet metric is re-instated and ignoring the dilation term which is suppressed by $1/k$, becomes
\EQ{
S_\text{eff}=-\frac k{8\pi}\int d^2x\,\STr\Big[
\CF ^{-1}\partial_\mu\CF\big(\sqrt{-h}h^{\mu\nu}G+\epsilon^{\mu\nu}B\big)\CF^{-1}\partial_\nu\CF\Big]\ .
%\label{ge3}
}
Then, the non-vanishing components are
\EQ{
T_{\pm\pm}&=\frac{\delta S_{eff}}{\delta h^{\pm\pm}}\Big|_{h_{\mu\nu}=\eta_{\mu\nu}}= -\frac{k}{8\pi} \STr(\CF^{-1}\partial_\pm \CF G \CF^{-1}\partial_\pm \CF)\\
&=-\frac{k}{8\pi} \STr\Big(A_\pm   
\big(\Omega\Omega^T-1\big)A_\pm \Big) = -\frac{k(1-\lambda^4)}{8\pi\lambda^4} \STr\Big(A_\pm ^{(2)}  
A_\pm ^{(2)}\Big)\,,
\label{emt}
}
where we have used \eqref{icr} to write it in terms of the gauge field and the two expressions for the metric $G$ in \eqref{metric}.
The surprising thing here is that they only depend on $A_\pm^{(2)}$ and not on the fermionic components. This dovetails with the fact that once the gauge field has been integrated out, the fermions only couple through a WZ term just like in the Green-Schwarz sigma model \eqref{pss}.

After imposing the Virasoro constraints the bosonic dimension is now 8, but the fermionic dimension is still 16 so we need a mechanism to reduce the number of physical fermions. The solution for the Green-Schwarz sigma model is the presence of fermionic gauge symmetries called kappa-symmetries. In favourable circumstance these can have the effect of halving the number of fermions to provide a perfect 8/8 matching.

The pertinent question is whether the deformed theories have kappa-symmetries.
The undeformed theory with $\Omega=1$ is an $F/F_V$ gauged WZW theory (with vector gauging) and this will give us a clue to uncovering the kappa-symmetries. A gauged $F/F_V$ WZW theory is {\it almost\/} invariant under local $F_L\times F_R$ symmetries. Namely, under the infinitesimal transformations
\EQ{
\delta\CF=\alpha\CF-\CF\beta\ ,\qquad \delta A_+=[\alpha, A_+] -\partial_+\alpha\ ,\qquad
 \delta A_-=[\beta, A_-] -\partial_-\beta\ ,
}
the transformation of the action is
\EQ{
\delta S= -\frac{k}{\pi}\int_\Sigma d^2x\, \STr\big(
(\alpha-\beta) F_{+-}  \big)\,,\qquad F_{+-}=\partial_+A_- -\partial_-A_+ +[A_+,A_-]\,.
\label{u8y}
}
Hence, invariance is ensured if $\alpha=\beta$, corresponding to vector gauging.\footnote{For $F/G$ with $G$ abelian, it is also possible to write a gauged WZW action invariant under axial gauge transformations simply by changing the sign of the term proportional to $\STr(A_+A_-)$ in~\eqref{gWZW}.}

The deformed theories~\eqref{dWZW} are obtained by replacing $\STr(A_+A_-)$ by $\STr(A_+\Omega A_-)$ in~\eqref{gWZW} and, therefore, \eqref{u8y} is replaced by
\EQ{
&\delta S=-\frac{k}{\pi}\int_\Sigma d^2x\, \STr\Big( (\Omega^T\alpha-\beta)\partial_+A_- -(\alpha-\Omega\beta)\partial_-A_+\\[5pt]
&\hspace{4.5cm} +\alpha[A_+,\Omega A_-] -\beta[\Omega^T A_+,A_-]\Big)\,.
}
In order to cancel the derivative terms, we take
\EQ{
\beta=\Omega^T\alpha\ ,\qquad \alpha=\Omega\beta\;\; \implies \;\; {\mathbb P}^{(2)}\alpha=0\,,
\label{exclude}
}
so that we are left with
\EQ{
\delta S=-\frac{k}{\pi}\int_\Sigma d^2x\, \STr\Big(\alpha[A_+,\Omega A_-] -\Omega^T\alpha[\Omega^T A_+,A_-]\Big)\,.
}
Clearly, $\delta S$ cancels if $\alpha\in\mf^{(0)}$, which corresponds to the $G$ (vector) gauge symmetry of the deformed theory.

However, we want to investigate the possibility of having
$\alpha\in\mf^{(1)}$ or $\mf^{(3)}$, which is allowed by~\eqref{exclude}. Taking the former case first, we get
\EQ{
\delta S=-\frac{k(1-\lambda^4)}
{\pi\lambda^2}\int d^2x\,\STr&\big(\alpha[A_+^{(1)},A_-^{(2)}]\big)
}
where, again, we have used~\eqref{couplings}. The way to cancel this term is familiar from the discussion of kappa-symmetry in \cite{Grigoriev:2007bu} (see also \cite{Arutyunov:2009ga}) so we shall not include all the details. The idea is to write the parameter of the variation as an anti-commutator
\EQ{
\alpha=[A_-^{(2)},\tilde\alpha]_+\ ,\qquad\tilde\alpha\in\mf^{(1)}\ .
\label{kappaL}
}
The fact that $\alpha$ and $\tilde\alpha$ are both elements of $\mf^{(1)}$ is checked in 
\cite{Arutyunov:2009ga}. Then, when the reality conditions are taken into account, the resulting number of independent symmetries is~8.

It then follows that
\EQ{
\delta S=-\frac{k(1-\lambda^4)}{
\pi\lambda^2}\int d^2x\,\STr\big(A_-^{(2)}A_-^{(2)}[\tilde\alpha,A_+^{(1)}]\big)\ .
}
Grigoriev and Tseytlin \cite{Grigoriev:2007bu} 
prove (in their Appendix D) that the above can be written as
\EQ{
\delta S&=-\frac{k(1-\lambda^4)}
{8\pi\lambda^2}\int d^2x\,\STr\big(A_-^{(2)}A_-^{(2)}\big)\STr\big(W[\tilde\alpha,A_+^{(1)}]\big)\\
&=\lambda^2\int d^2x\,\STr\big(W[\tilde\alpha,A_+^{(1)}]\big)T_{--}\ ,
\label{www}
}
where, in the defining representation, $W=\text{diag}(1,1,1,1,-1,-1,-1,-1)$ is the fermionic parity operator, or hypercharge. This can be shown simply by writing $A_-^{(2)}$ as
\EQ{
A_-^{(2)}= g_-(c_1 \Lambda_1 + c_2\Lambda_2) g_-^{-1}\,,\qquad g_-\in G\,,
}
with
\EQ{
\Lambda_1=\frac{i}2 \text{diag}(t,0)\,,\qquad\Lambda_2=\frac{i}2 \text{diag}(0,t)\,,\qquad t=\text{diag}(1,1,-1,-1)\,.
\label{prelambda}
}
This is a consequence of the, so called, ``polar coordinate decomposition'' applied to $F_B/G$, which is a product of ordinary symmetric spaces of rank~1.\footnote{Notice that there are field configurations $A_-^{(2)}$ of the same form with $\STr(\Lambda_1^2)<0$ or $\STr(\Lambda_1^2)=0$~\cite{Miramontes:2008wt}, but they are not directly related to the world-sheet theory of the string in $\text{AdS}_5\times S^5$ \cite{Grigoriev:2007bu,Hoare:2012nx}. The reason is that, for those configurations, the Virasoro constraint $T_{--}=0$ leads to either $A_-^{(2)}=0$ or $A_-^{(2)}\in {\mathfrak so}(1,4)^{(2)}$, respectively. In contrast, for~\eqref{prelambda} it leads to $c_1=c_2$, and $A_-^{(2)}$ has non-trivial components in both ${\mathfrak so}(1,4)^{(2)}$ and ${\mathfrak so}(5)^{(2)}$.}
Then,
\EQ{
A_-^{(2)}A_-^{(2)}=\frac{1}8 \STr(A_-^{(2)}A_-^{(2)}) W + \frac{1}8(c_1^2+c_2^2) {\mathbb I}_{8}\,,
\label{nice}
}
which leads to the identity used in~\eqref{www}.

In conformal gauge, one imposes the Virasoro constraint $T_{--}=0$ directly, in which case \eqref{www} vanishes, or if the worldsheet metric is not fixed then the
variation $\delta S$ in \eqref{www} being proportional to $T_{--}$ can be cancelled by a suitable variation of the world-sheet metric involving a parameter proportional to $\STr\big(W[\tilde\alpha,A_+^{(1)}]\big)$.

There is a similar discussion for $\alpha\in\mf^{(3)}$; in this case
\EQ{
\delta S=\frac{k(1-\lambda^4)}
{\pi\lambda^3}\int d^2x\,\STr&\big(\alpha[A_+^{(2)},A_-^{(3)}]\big)
}
which, by choosing
\EQ{
\alpha=[A_+^{(2)},\hat\alpha]_+\,,\qquad \hat\alpha\in {\mathfrak f}^{(3)}\,,
}
becomes
\EQ{
\delta S=-\frac{k(1-\lambda^4)}
{\pi\lambda^3}\int d^2x\,\STr&\big(A_+^{(2)}A_+^{(2)} [\hat{\alpha}, A_-^{(3)}]\big)=
\lambda\int d^2x\,\STr\big(W[\hat\alpha,A_+^{(1)}]\big)T_{++}\,.
}
Again, this gives rise to 8 independent symmetries of this type.

We can confirm the resulting number of independent kappa-symmetries as follows. Imposing the Virasoro constraint $T_{--}=0$, the field $A_-^{(2)}$ is of the form
\EQ{
A_-^{(2)}= c_1g_-(\Lambda_1+\Lambda_2)g_-^{-1}\equiv c_1g_-\Lambda g_-^{-1}\,,\qquad g_-\in G\,.
}
Then, the algebra $\mf$ has a well-defined orthogonal decomposition into the kernel and the image of $A_-^{(2)}$~\cite{Grigoriev:2007bu}
\EQ{
\mf=\mf_-^\perp\oplus\mf_-^\parallel\ ,\qquad
[A_-^{(2)},\mf_-^\perp]=0\ ,\qquad \mf_-^\parallel=[A_-^{(2)},\mf]\ .
\label{ppz}
}
Projectors onto $\mf_-^\perp$ and $\mf_-^\parallel$ are defined using the anti-commutator and commutator with $A_-^{(2)}$, respectively:
\EQ{
\mf_-^\perp=[A_-^{(2)},\mf]_+\ ,\qquad \mf_-^\parallel=[A_-^{(2)},\mf]\ .
\label{pjr}
}
Therefore, the number of independent parameters of the form $[A_-^{(2)},\tilde{\alpha}]_+$ equals the dimension of $\mf_-^\perp$.
Since for $\text{PSU}(2,2|4)$ the dimensions are generically
\EQ{
\DF^{(1)\parallel}=\DF^{(1)\perp}=\DF^{(3)\parallel}=\DF^{(3)\perp}=8\ .
}
the resulting number of independent symmetries of this type is indeed~8.

In conclusion, these kappa-symmetries, reduce the number of fermionic degrees-of-freedom from 16 to 8 to match the number of bosons. To summarize:

\greybox{\begin{center}\underline{\bf The symmetries}\end{center}
\noindent
{\bf Gauge symmetry:}
\EQ{
\delta\CF&=[\epsilon^{(0)},\CF]\ ,\\ \delta A_\pm&=[\epsilon^{(0)},A_\pm]-\partial_\pm\epsilon^{(0)}\ .
\label{ps1}
}

\noindent
{\bf Kappa-symmetries:}
\EQ{
\delta\CF&= \epsilon^{(1)}\CF-\lambda\CF\epsilon^{(1)}\ ,\\
\delta A_+&=[\epsilon^{(1)},A_+]-\partial_+\epsilon^{(1)}\ ,\\
\delta A_-&=\lambda[\epsilon^{(1)},A_-]-\lambda\partial_-\epsilon^{(1)}\ ,
\label{ps2}
}
for $\epsilon^{(1)}=[A_-^{(2)},\kappa^{(1)}]_+$; and
\EQ{
\delta\CF&= \lambda\epsilon^{(3)}\CF-\CF\epsilon^{(3)}\ ,\\
\delta A_+&=\lambda[\epsilon^{(3)},A_+]-\lambda\partial_+\epsilon^{(3)}\ ,\\
\delta A_-&=[\epsilon^{(3)},A_-]-\partial_-\epsilon^{(3)}\ ,
\label{ps3}
}
for $\epsilon^{(3)}=[A_+^{(2)},\kappa^{(3)}]_+$. In the above, $\kappa^{(i)},\epsilon^{(i)}\in\mf^{(i)}$.
}

\section{Discussion}\label{s6}

We have investigated a way to deform the world-sheet theory for the string on AdS$_5\times S^5$ whilst preserving integrability. The deformed theory has kappa-symmetries that are needed to get the right count of the fermionic degrees-of-freedom. When written in terms of appropriate variables, the deformed theories have the same equations-of-motion as the Green-Schwarz sigma model for AdS$_5\times S^5$ but a deformed symplectic structure. 

We conjecture that the deformed theories provide a Green-Schwarz Lagrangian formulation of the deformed S-matrices constructed in \cite{Hoare:2011nd,Hoare:2011wr,Hoare:2012fc,Hoare:2013ysa}. This conjecture implies that the deformation parameter $\lambda$ is an exactly marginal deformation of the deformed theories which is also required if the deformed theories define consistent world-sheet theories with a vanishing beta function. It would interesting, as mentioned in section \ref{s3.5} to calculate the beta function to one loop.

Naturally it would be interesting to investigate the deformed theories from a target spacetime perspective. This will require extracting the spacetime fields and verifying that the generalized Einstein equations are satisfied. 

In a companion paper \cite{two}, we will undertake a Hamiltonian analysis of the deformed theory and show how in the $\lambda\to0$ limit, the deformed worldsheet theory becomes the 
semi-symmetric space sine-Gordon theory originally constructed  in \cite{Grigoriev:2007bu,Mikhailov:2007xr} and studied in detail in \cite{Schmidtt:2010bi,Goykhman:2011mq,Hollowood:2011fq,Schmidtt:2011nr}. These generalized sine-Gordon theories build on the original bosonic theories in \cite{Park:1994bx,Hollowood:1994vx,Bakas:1995bm,FernandezPousa:1996hi}.
The semi-symmetric sine-Gordon theory is a relativistic integrable theory which has an ${\cal N}=(8,8)$ supersymmetry defined by the action
\SP{
S&=S_\text{gWZW}[\gamma,W_\mu]-\frac{k\mu^2}{\pi}\int d^2x\,\text{STr}\,\big(\Lambda
\gamma^{-1}\Lambda\gamma\big)\\
&+\frac k{2\pi}\int d^2x\,\text{STr}\,\big(\frac1\mu\,\psi_+[\Lambda,D_-\psi_+]-\frac1\mu\,\psi_-[\Lambda,D_+\psi_-]
-2\psi_+\gamma^{-1}\psi_-\gamma\big)\ ,
\label{alp}
}
where $S_\text{gWZW}[\gamma,W_\mu]$ is the action of the $G/H$ gauged WZW model  with level $k$. Here $\Lambda=\Lambda_1+\Lambda_2=\frac{i}2\,\text{diag}(t,t)$ and $H$ is the centralizer of $\Lambda$ in $G$. It is important to realize that the potential and fermionic terms in the above are written in the Lie superalgebra of $F$ rather than $G$. These theories have a non-trivial vacuum structure and admit a rich spectrum of topological solitons or kinks \cite{Hollowood:2009tw,Hollowood:2010rv,Hollowood:2010dt,Hollowood:2011fm,Hollowood:2013oca}.

\vspace{0.2cm}
\begin{center}
{\tiny******}
\end{center}
\vspace{0.2cm}

\noindent We would like to thank Kostas Sfetsos for fruitful discussions. TJH is supported in part by the STFC grant ST/G000506/1. 
JLM is supported in part by MINECO (FPA2011-22594), the Spanish Consolider-
Ingenio 2010 Programme CPAN (CSD2007-00042), Xunta de Galicia (GRC2013-024),
and FEDER.
DMS is supported by the FAPESP-BEPE grant 2013/23328-1.

\appendix
\appendixpage

\section{Lie Superalgebras}\label{a1}

Our focus is on the semi-symmetric space that underlies the Green-Schwarz sigma model for superstrings in AdS$_5\times S^5$.
Our conventions for superalgebras are  
taken from the review \cite{Arutyunov:2009ga} (see also~\cite{Kac:1977em}). The superalgebra ${\mathfrak sl}(4|4)$ is defined by the $8\times8$ matrices
\EQ{
M=\MAT{m&\theta\\ \eta&n}\ ,
\label{mat}
}
where $m$ and $n$ are Grassmann even and $\theta$ and $\eta$ are Grassmann odd. 
These matrices are required to have vanishing supertrace\footnote{Notice that our convention is the opposite of \cite{Grigoriev:2007bu}, so that the supertrace is positive on the $S^5$ factor and negative on the $AdS_5$ factor.}
\EQ{
\STr M=-\Tr\,m+\Tr\,n=0\ .
\label{STr}
}
The non-compact real form ${\mathfrak su}(2,2|4)$ is picked out by imposing the reality condition
\EQ{
M=-HM^\dagger H
\label{real}
}
where, in $2\times 2$ block form,
\EQ{
H=\left(\begin{array}{cc|cc} {\mathbb I}_2 &&&\\ &-{\mathbb I}_2 &&\\ \hline &&{\mathbb I}_2&
\\ &&&{\mathbb I}_2\end{array}\right)\,.
}
Here, $\dagger$ is the usual hermitian conjugation, $M^\dagger=(M^*)^t$, but with the definition that complex conjugation is anti-linear on products of Grassmann odd elements
\EQ{
(\theta_1\theta_2)^*=\theta_2^*\theta_1^*\ ,
}
which guarantees that $(M_1M_2)^\dagger=M_2^\dagger M_1^\dagger$.
The superalgebra ${\mathfrak{psu}}(2,2|4)$ is then the quotient of $\mathfrak{su}(2,2|4)$ by the unit element $i{\mathbb I}_{8}$, which is a centre of the algebra. 

An important r\^ole is played by a ${\mathbb Z}_4$ autormorphism defined as
\EQ{
M\longrightarrow\sigma_-(M)=-{\cal K}M^{st}{\cal K}^{-1}\ ,
}
where $st$ denotes the ``super-transpose'' defined as
\EQ{
M^{st}=\MAT{m^t&-\eta^t\\ \theta^t&n^t}\ .
}
In the above,
\EQ{
{\cal K}=\left(\begin{array}{cc|cc} J_2 &  &  & \\
  & J_2 &  & \\ \hline
 &  & J_2 & \\
 &  & & J_2\end{array}\right)\ ,\qquad J_2=\MAT{0&-1\\ 1&0}\ .
}
Under $\sigma_-$, the superalgebra ${\mathfrak{psu}}(2,2|4)$ has the decomposition
\EQ{
{\mathfrak f}= {\mathfrak f}^{(0)}\oplus{\mathfrak f}^{(1)}\oplus{\mathfrak f}^{(2)}\oplus{\mathfrak f}^{(3)}\ ,\qquad \sigma_-({\mathfrak f}^{(j)})=i^j\,{\mathfrak f}^{(j)}\ , \qquad [{\mathfrak f}^{(j)},{\mathfrak f}^{(k)}]\subset {\mathfrak f}^{(j+k\; \text{mod}\; 4)}\,.
\label{CanonicalDecN}
}
In particular, the even graded parts are Grassmann even while the odd graded parts are Grassmann odd. The zero graded part ${\mathfrak f}^{(0)}\equiv{\mathfrak g}$ is the (bosonic) Lie algebra
of $G$, which is the group in the denominator of the semi-symmetric space.

Currents and gauge fields take values in $\mathfrak{psu}(2,2|4)$, and it is convenient to write a matrix  like~\eqref{mat} in terms of a basis of ordinary $8\time 8$ matrices which respects the ${\mathbb Z}_4$ grading
\EQ{
M= \sum_{i=0}^{3} m_{a_i} T^{a_i}\,,\qquad T^{a_i}\in {\mathfrak f}^{(i)}\,.
}
The coefficients $m_{a_0}$ and $m_{a_2}$ are Grassmann even, while $m_{a_1}$ and $m_{a_3}$ are Grassmann odd. In more precise terms, $T^a$ are the generators of ${\mathfrak f}=\mathfrak{psu}(2,2|4)$ and this construction defines the so-called Grassmann envelope of ${\mathfrak f}$. Then, any two matrices satisfy
\EQ{
[M,N]= m_{a_i} n_{a_j} (T^{a_i}T^{a_j}- (-1)^{ij} T^{a_j}T^{a_i})\equiv m_{a_i} n_{a_j} [T^{a_i}, T^{a_j}\}=-[N,M]\,,
}
where $[\cdot,\cdot\}$ is the superbracket. In particular this ensures that
\EQ{
\STr([M,N])=0\,.
}

The supertrace defines a bilinear form on the generators $T^a$
\EQ{
\STr(T^{a} T^{b})= \eta^{ab}=(-1)^{[a][b]} \eta^{ba}\,.
}
It is invariant,
\EQ{
\STr([T^a,T^b\}T^c)=\STr(T^a[T^b,T^c\})\,.
}
Moreover, since we will always take a basis of generators that respects the  ${\mathbb Z}_4$ grading, the only non-vanishing components are $\eta^{a_0b_0}$, $\eta^{a_2b_2}$ and $\eta^{a_1b_3}=-\eta^{b_3a_1}$.

\end{document}